\documentclass[aps,prl,twocolumn,superscriptaddress]{revtex4}

\usepackage{graphicx}
\usepackage{color}

\begin{document}

\title{Ramifications of Optical Pumping on the Interpretation of Time-Resolved Photoemission Experiments on Graphene}

\author{S\o ren Ulstrup}
\affiliation{Department of Physics and Astronomy, Interdisciplinary Nanoscience Center (iNANO), Aarhus University, Denmark}
\author{Jens Christian Johannsen}
\affiliation{Institute of Condensed Matter Physics, \'Ecole Polytechnique F\'ed\'erale de Lausanne (EPFL), Switzerland}
\author{Federico Cilento}
\author{Alberto Crepaldi}
\affiliation{Elettra - Sincrotrone Trieste S.C.p.A., 34149 Basovizza, Trieste, Italy}
\author{Jill A. Miwa}
\affiliation{Department of Physics and Astronomy, Interdisciplinary Nanoscience Center (iNANO), Aarhus University, Denmark}
\author{Michele Zacchigna}
\affiliation{IOM-CNR Laboratorio TASC, Area Science Park, Trieste, Italy}
\author{Cephise Cacho}
\author{Richard T. Chapman}
\affiliation{Central Laser Facility, STFC Rutherford Appleton Laboratory, Harwell, United Kingdom}
\author{Emma Springate}
\affiliation{Central Laser Facility, STFC Rutherford Appleton Laboratory, Harwell, United Kingdom}
\author{Felix Fromm}
\author{Christian Raidel}
\author{Thomas Seyller}
\affiliation{Institut f{\"u}r Physik, Technische Universit{\"a}t Chemnitz , Germany}
\author{Phil D. C. King}
\affiliation{SUPA, School of Physics and Astronomy, University of St. Andrews,
St. Andrews, United Kingdom}
\author{Fulvio Parmigiani}
\affiliation{Elettra - Sincrotrone Trieste S.C.p.A., 34149 Basovizza, Trieste, Italy}
\affiliation{Department of Physics, Universit{\'a} degli Studi di Trieste, 34127 Trieste, Italy}
\affiliation{International Faculty - University of K{\"o}ln, Germany}
\author{Marco Grioni}
\affiliation{Institute of Condensed Matter Physics, \'Ecole Polytechnique F\'ed\'erale de Lausanne (EPFL), Switzerland}
\author{Philip Hofmann}
\affiliation{Department of Physics and Astronomy, Interdisciplinary Nanoscience Center (iNANO), Aarhus University, Denmark}
\email{philip@phys.au.dk}


\begin{abstract}
In pump-probe time and angle-resolved photoemission spectroscopy (TR-ARPES) experiments the presence of the pump pulse adds a new level of complexity to the photoemission process in comparison to conventional ARPES. This is evidenced by pump-induced vacuum space-charge effects and surface photovoltages, as well as multiple pump excitations due to internal reflections in the sample-substrate system. These processes can severely affect a correct interpretation of the data by masking the out-of-equilibrium electron dynamics intrinsic to the sample. In this study, we show that such effects indeed influence TR-ARPES data of graphene on a silicon carbide (SiC) substrate. In particular, we find a time- and laser fluence-dependent spectral shift and broadening of the acquired spectra, and unambiguously show the presence of a double pump excitation. The dynamics of these effects is slower than the electron dynamics in the graphene sample, thereby permitting us to deconvolve the signals in the time domain. Our results demonstrate that complex pump-related processes should always be considered in the experimental setup and data analysis.
\end{abstract}
\maketitle

\section{Introduction}
\begin{figure*}
\includegraphics[width=0.9\textwidth]{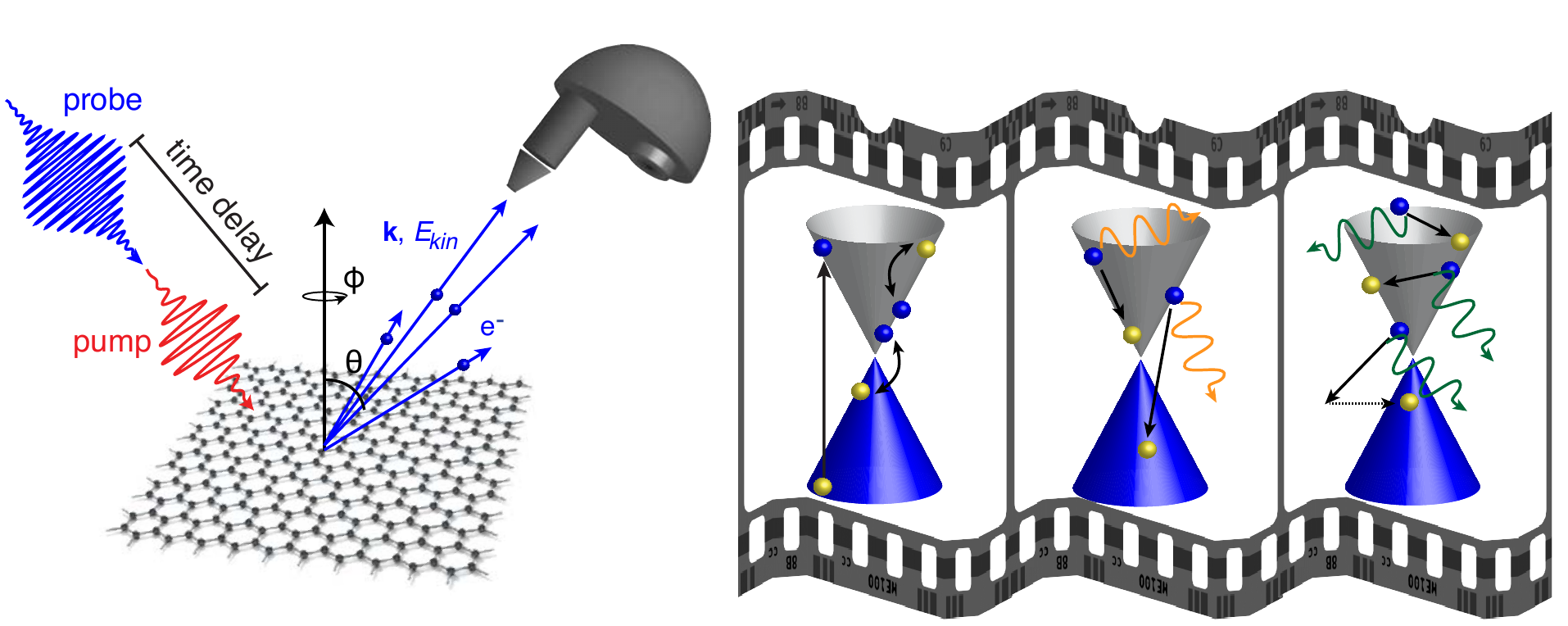}\\
\caption{Schematic of a TR-ARPES experiment on graphene. Electrons (blue spheres) excited by an infrared pump pulse are photoemitted by an ultraviolet probe on an ultrashort timescale and detected by an electron analyzer. This provides energy-, momentum- and time-resolved snapshots of the electron dynamics around the Dirac cone. Such measurements provide direct information about electron and hole (yellow spheres) recombination via electron-electron (curled arrows in left Dirac cone sketch) and electron-phonon (wiggled arrows) processes. Interactions between electrons and optical phonons (middle Dirac cone sketch), acoustic phonons and impurities (right Dirac cone sketch) can be disentangled by the timescales revealed in the non-equilibrium signal.}
\label{fig:1}
\end{figure*}
Angle-resolved photoemission spectroscopy (ARPES) offers a unique capability to investigate the electronic properties of quasiparticles in complex systems with energy and momentum resolution. Recent developments have seen significant advancements in adding femtosecond time resolution to ARPES in a pump-probe scheme \cite{Perfetti:2007,Schmitt:2008,Petersen:2011,Rohwer:2011}. New insights into the nature and origin of intriguing physical phenomena, such as collective excitations and phase transitions, can be gained by observing the time-dependent changes directly in the band structure. More specifically, time-resolved ARPES (TR-ARPES) gives access to the time-dependent changes of the spectral weight in selected states in the Brillouin zone. The different time scales observed connect to the different energy scales characteristic of the various kinds of interactions simultaneously present in the material \cite{Rohwer:2011}. Consequently, interactions felt by the quasiparticles in specific states can be disentangled by their characteristic time scales. This time, energy and momentum resolving capability is a virtue precluded to time resolved all-optical techniques. It has opened unprecedented possibilities to study the quasiparticle dynamics in a wide range of condensed matter systems, including the ultrafast amplitude oscillations in charge density wave materials \cite{Schmitt:2008,Rohwer:2011,Petersen:2011,Hellmann:2010}, the quasiparticle recombination rate near the \emph{d}-wave node in cuprate superconductors \cite{Perfetti:2007,Cortes:2011,Graf:2011aa,Smallwood:2012}, the dynamics taking place on the Dirac cone in graphene \cite{Johannsen:2013aa,Ulstrup:2014b,Gierz:2013aa,Johannsen:2015,Ulstrup:2015a} and on the surface of topological insulators \cite{Crepaldi:2012,crepaldi2013,Sobota:2012,Wang:2012d,Wang:2013c,Hajlaoui:2014aa}. 

A TR-ARPES experiment is performed by first exciting the system out of equilibrium by electron-hole pair creation using a femtosecond pump pulse with an energy below the sample work function. After a variable time delay, the excited electron distribution is photoemitted using a second probe pulse with an energy above the work function threshold, as illustrated in Fig.~\ref{fig:1} for the case of monolayer graphene. As the time delay is varied, the acquired ARPES spectra monitor the relaxation of this excited distribution back to the equilibrium state. This relaxation process involves a complex interplay of ultrafast electron-electron and electron-phonon scattering events. For monolayer graphene on SiC, several recent studies report that the initial thermalization of the excited carriers to a hot Fermi-Dirac (FD) distribution can be accompanied by a multiplication of the number of hot electrons with energies larger than the Fermi energy \cite{Rana:2007,Winzer:2010,Song:2013,Brida:2013aa,Tielrooij:2012,Ploetzing:2014,Johannsen:2015}, and that the subsequent cooling of the hot electron distribution happens on different time scales determined by energy relaxation via optical phonons and acoustic phonons, primarily assisted by disorder-scattering (so-called super collision processes) \cite{Song:2012c,Johannsen:2013aa,Bistritzer:2009,Butscher:2007}. These processes are sketched in Fig.~\ref{fig:1} on the Dirac cones in movie snapshots.

To extract this intrinsic electron dynamics from the acquired TR-ARPES data, great care should be exercised as several experimental effects might influence the momentum and energy distributions of the photoelectrons before, during and after photoemission. In fact, in a low repetition rate laser source ($\leq1$~kHz) the intensity of both pump and probe pulses is necessarily high in order to detect a pump-probe signal out of the noise within a reasonable experimental time window. This implies a high electron density in the cloud of photo-emitted electrons propagating towards the electron analyzer. The mutual Coulomb interactions between the photoelectrons create a vacuum space-charge effect that may lead to a shift of the electrons' kinetic energies and a severe broadening in energy that can easily exceed the resolution set by the light source and the analyzer \cite{Zhou:2005, Hell:2009,Passlack:2006}. Moreover, a very recent time-resolved x-ray photoelectron spectroscopy study by L.P. Oloff \emph{et al.} \cite{Oloff:2014} shows that photoelectrons emitted by the x-ray probe beam may also be influenced by the electric field originating from secondary electrons generated by the infrared pump. This may appear counterintuitive as the pump energy is typically lower than the sample work function ($< 4$~eV). However, high-order nonlinear photoemission processes, where an electron simultaneously absorbs several photons, can provide sufficient energy for the electrons to escape into vacuum \cite{Oloff:2014,Damascelli:1996}. These processes can dominate as the pump power ($\approx 1$~mW) is typically several orders of magnitude higher than the probe power ($\approx 1$~nW). For semiconductors, an additional effect, originating from a light-induced voltage difference in the surface region, can affect the propagating photoelectrons. Specifically, at the surface, the photo-excited electron-hole pairs can be spatially separated by the band bending. This separation generates a transient electric field that affects the photoelectron outside the sample. Despite having been known for several years, it was only very recently that the implications of this surface photovoltage (SPV) for TR-ARPES experiments were addressed theoretically by S. Tanaka \cite{Tanaka:2012a} and experimentally by S.-L. Yang \emph{et al.} on semiconducting GaAs \cite{Yang:2014aa}.  

In this paper, we investigate the extent of these perturbing effects in the acquisition of TR-ARPES data from monolayer graphene on a semiconducting SiC substrate. We find strong indications of pump-induced electric fields and pump beam reflections at the backside of the SiC substrate that affect the measurement on picosecond timescales. A thorough understanding of these effects is important in order to suppress their impact on the experiment or deconvolve them from the intrinsic electron dynamics in the data analysis.

\section{Experiment}
The graphene sample investigated in this work was produced following a well-documented synthesis method that provides high-quality monolayer graphene on the Si-terminated face of SiC(0001). The graphene layer was decoupled from the substrate by hydrogen intercalation \cite{Riedl:2009,Speck:2011} such that the structural and electronic properties probed by transport, Raman and ARPES measurements closely approached pristine graphene \cite{Barreto:2013,Johannsen:2013,Bostwick:2010,Speck:2011}. The graphene is $p$-doped with a carrier concentration of $\approx$~5~$\times~10^{12}$~cm$^{-2}$ that places the Dirac point 240 meV above the Fermi energy. The doping mechanism is caused by a spontaneous polarization in the bulk of the semiconducting SiC substrate \cite{Ristein:2012}. The sample was cleaned by annealing to 500 K in ultra-high vacuum in order to remove adsorbed water, and was held at room temperature throughout the entire experiment. The thickness of the SiC wafer was $390$~$\mu$m.  

All TR-ARPES spectra presented in this work were acquired at the Artemis facility, Rutherford Appleton Laboratory using the TR-ARPES end-station \cite{Cacho:2014, Frassetto:2011}. The extreme ultraviolet, ultrafast probe pulses provided at this facility by high-harmonic generation enable acquisition of TR-ARPES data from the $\bar{K}$-point at the corner of the hexagonal Brillouin zone of graphene. The sample was excited by using a beam of 30~fs light pulses with a repetition rate of 1~kHz, a photon energy of $h\nu=$ 1.55~eV and with fluences ranging from $1$~mJ/cm$^2$ to $6.6$~mJ/cm$^2$. The band structure around the Dirac cone was measured by a time delayed probe pulse of $h\nu=$ 21~eV corresponding to the 13th harmonic of the laser fundamental. For the measurements in this work, a large time delay window spanning from $-100$~ps to $100$~ps was used in all scans. Negative time delays correspond to a situation where the probe pulse arrives before the pump pulse. The grid of time delay points was varied in such a way that a coarse grid was used for long time delays, while a fine grid was used to properly capture the hot electron signal around the time of the optical excitation that we define as time zero ($t=0$). The time and angular resolutions were better than 40~fs and 0.3~$^{\circ}$, respectively. The intrinsic energy resolution was 380~meV in these experiments.  Additional fluence dependent broadening effects were present, leading to an apparent energy resolution denoted as $\Delta E$, which we discuss in the following sections.

\section{Results and Discussion}

\subsection{Snapshots of electron dynamics}

\begin{figure*}[t!]
\includegraphics[width=1\textwidth]{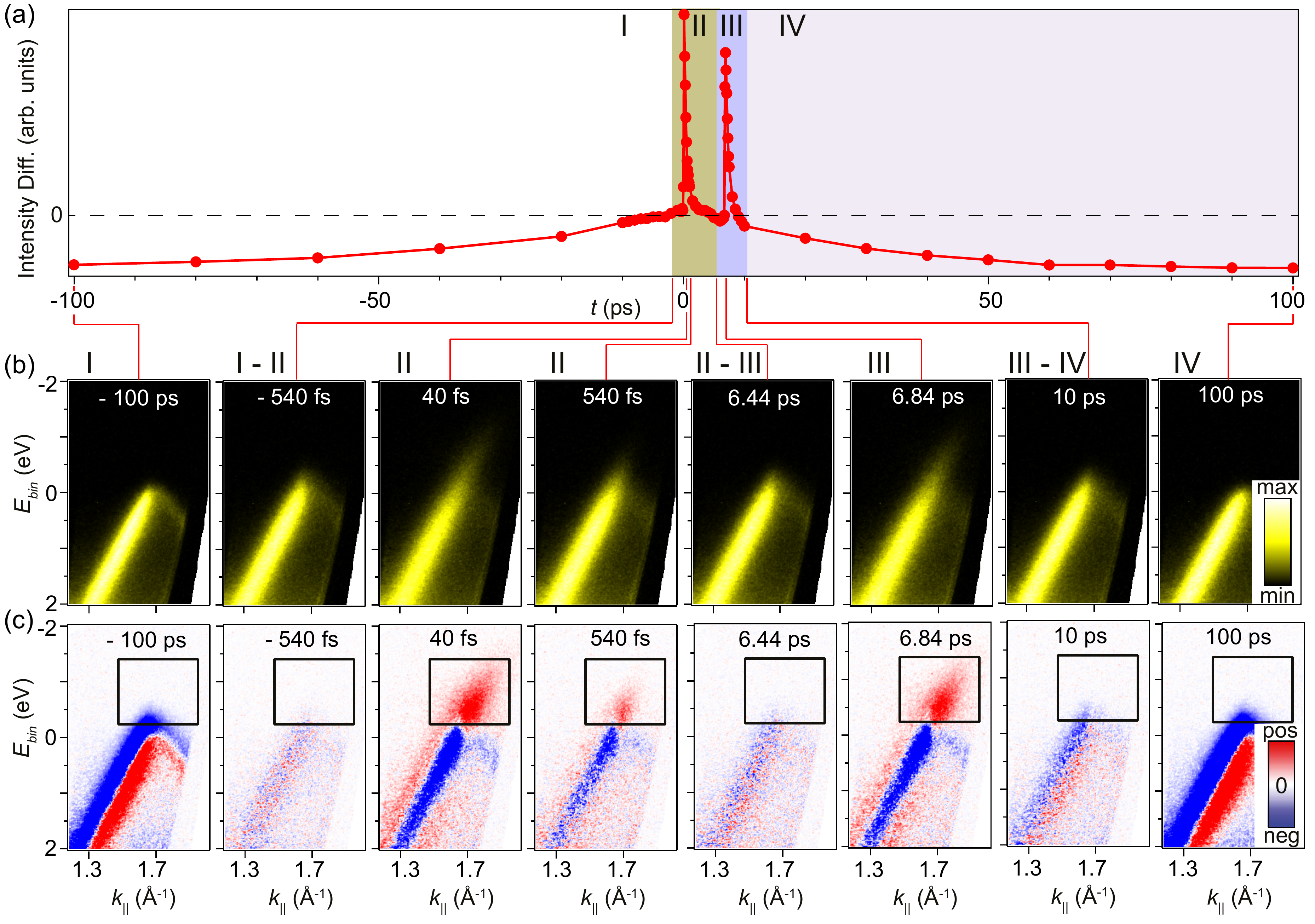}\\
\caption{Electron dynamics in graphene induced by a pump laser pulse with an energy of 1.55 eV and a fluence of $2.2$~mJ/cm$^2$ over a $\pm100$~ps time window: (a) Time dependent intensity difference summed within the boxed region given in the intensity difference plots in (c). The dynamics has been divided into regions I, II, III and IV with representative TR-ARPES spectra shown in (b) for each region and for the time delays defining the boundaries between the regions. (c) Intensity difference obtained by subtracting a spectrum averaged over the time interval from $-1000$~fs to $-100$~fs from the spectrum at the given time delay.}
\label{fig:2}
\end{figure*}

TR-ARPES measurements of the low-energy Dirac spectrum of graphene at a pump fluence of $2.2$~mJ/cm$^2$ are shown in Fig. \ref{fig:2}. The acquired dynamics is divided into four temporal regions labeled I, II, III and IV that are marked in Fig. \ref{fig:2}(a). Snapshots of the Dirac cone of graphene at selected time delays are presented in Fig. \ref{fig:2}(b), and corresponding intensity difference plots are shown in Fig. \ref{fig:2}(c). The latter are arbitrarily determined by averaging over all spectra taken in the time interval from $-1000$~fs to $-100$~fs and subtracting this average from the spectrum at any given time delay. The Fermi level is referenced to the spectrum at -100~ps. The intensity difference in Fig. \ref{fig:2}(a) was integrated over the boxed region in the intensity difference plots shown in Fig. \ref{fig:2}(c). Time zero is fixed on the middle of the rising edge of the initial optical excitation (in region II) of the electrons in the sample.

Several unexpected features emerge in the dynamics of the intensity difference in Fig. \ref{fig:2}(a). Perhaps most striking is the presence of two large peaks (one in region II and one in region III) both originating from a pump excitation. The first occurs as the pump beam impinges on the surface and the second is due to a reflection of the pump beam at the back of the substrate. We present a more detailed discussion of this below. Furthermore, in the temporal regions I and IV a depletion of intensity difference with respect to the averaged spectrum immediately before the excitation is seen towards both large negative and large positive time delays. Comparing the spectrum at a time delay of $-100$~ps to the one at $-540$~fs (here defined as the boundary between regions I and II) in Fig. \ref{fig:2}(b), one observes that the spectrum at -100~ps is shifted rigidly towards higher binding energies and appears less broad. The intensity difference in Fig. \ref{fig:2}(c) at -100~ps reflects this behavior directly as there is a surplus (depletion) of intensity towards higher (lower) binding energies. In region IV it is found that the spectrum also shifts to higher binding energies with a small shift already evident around 10~ps. We can therefore define an apparent chemical potential $\mu$ and apparent energy resolution $\Delta E$ that both change with time. 

Optical excitation of electrons is observed in region II as a large peak, which rapidly decays on a femtosecond time scale and is fully relaxed after $\approx6$~ps. This behavior reflects hot electron dynamics in graphene i.e. thermal relaxation of hot electrons by phonon scattering. The additional pump excitation triggers this dynamics again in region III. The fact that the electronic system is heated twice is also evident from the smearing of the FD edge in the spectra and difference plots at both 40~fs and 6.84~ps in Figs. \ref{fig:2}(b)-(c). 

\subsection{Influence of pump-induced electric fields}

\begin{figure*} [t!]
\includegraphics[width=1\textwidth]{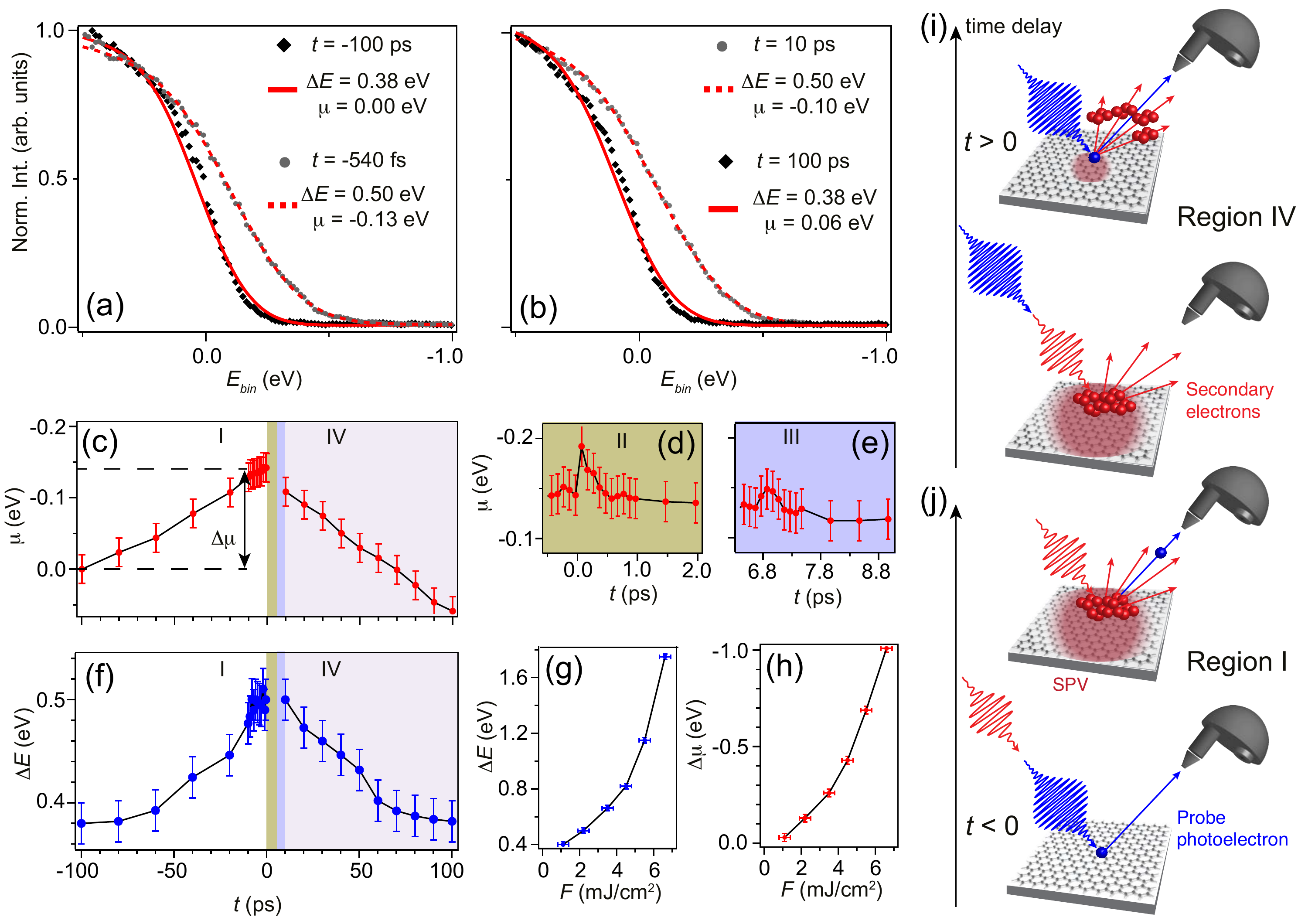}
\caption{(a)-(b) MDC-derived energy distribution curves (dots) at the given time delays obtained from MDC fits to the spectra and fitted to a FD function convoluted with a Gaussian (lines). The stated values for the apparent chemical potential $\mu$ and the apparent energy resolution $\Delta E$ are results of the fits. (c)-(e) Fitted position of $\mu$ for (c) long positive (region IV) and negative (region I) delay times and for (d)-(e) time-domain regions II and III where the pump-induced hot electron dynamics in graphene takes place. (f) Fitted values of $\Delta E$ as a function of time delay. The energy broadening is not defined in regions II and III where the electronic system is heated. The pump laser fluence in (a)-(f) is 2.2~mJ/cm$^2$. (g) Apparent energy resolution at a time delay of -200~fs and (h) chemical potential shift $\Delta\mu$ defined in (c) as a function of pump laser fluence. (i)-(j) Sketch of pump-probe scenarios at (i) long positive delays (region IV) and (j) long negative delays (region I). Blue spheres signify photoelectrons emitted by the probe and red spheres signify secondary electrons generated by the pump. The red disk is a surface photovoltage (SPV) induced in the SiC substrate by the pump pulse. The time it takes a photoelectron to be detected at the analyzer is much longer than the typical time delay between the pump and probe pulses.}
\label{fig:3}
\end{figure*}

In order to track more closely the evolution of $\mu$ and the change of $\Delta E$, we apply a procedure that accurately provides the energy distribution curves of the electrons at all time delays by analyzing momentum distribution curves (MDCs) of each spectrum as described in detail elsewhere \cite{ulstrupc:2014}. Examples of such MDC-derived energy distribution curves and the fits to a FD function convoluted with a Gaussian energy function with width $\Delta E$ are shown for several time delays in Figs. \ref{fig:3}(a)-(b). 

The complete time dependence of $\mu$ in regions I and IV is given in Fig. \ref{fig:3}(c). In region I the value of $\mu$ moves to lower binding energies until a plateau is reached around the time of the optical excitation. In region IV $\mu$ moves back to higher binding energies. A similar trend is observed for $\Delta E$ in Fig. \ref{fig:3}(f). In regions II and III the position of $\mu$ changes due to the hot electron dynamics intrinsic to the sample. The population of photoinduced electrons above the Dirac point forces the chemical potential towards lower binding energies as seen in the fits in Figs. \ref{fig:3}(d)-(e). This behavior is perfectly consistent with a simple calculation of $\mu$ for graphene with elevated electronic temperatures assuming that the charge density is conserved \cite{Johannsen:2015}.

The behavior of $\mu$ and $\Delta E$ in regions I and IV does not reflect the instrinsic electron dynamics of the graphene but is the result of pump-induced external electric fields that affect the kinetic energy of each photoelectron emitted by the XUV probe pulse as it propagates to the electron analyzer \cite{Zhou:2005, Hell:2009,Passlack:2006}. As mentioned above, such fields may originate from either a vacuum space-charge effect, a SPV effect in the SiC substrate \cite{Yang:2014aa} or a combination of both. The space-charge is caused by electrostatic interactions between the probe photoelectrons and secondary electrons excited by the pump (1.55 eV) to energies higher than the sample work function by nonlinear multiple photon absorption processes. These perturbing effects, which can persist on the nanosecond timescale, are particularly effective at the low repetition rate and high fluences we use, which yield large energies per pulse \cite{Damascelli:1996,Oloff:2014}. This type of nonlinear process can also lead to the creation of electron-hole pairs in SiC by exciting valence band electrons across the band gap. A SPV is then created, generating a dipole field that will also affect the kinetic energy of the probe photoelectrons \cite{Tanaka:2012a,Yang:2014aa}. The sketch in Fig. \ref{fig:3}(i) illustrates the event sequence for these effects at positive delay times ($t>0$) corresponding to the dynamics in region IV.   Initially the pump simultaneously induces a SPV and generates secondary electrons that propagate from the sample surface with low kinetic energies. When the probe arrives at a later time delay, it photoemits electrons which are decelerated as they move towards the secondary electron charge cloud. Due to the much higher kinetic energies of the photoelectrons emitted by the UV pulse, they pass through the charge cloud and are re-accelerated towards the electron analyzer. Photoelectrons emitted by the probe at a negative time delay ($t<0$) in region I of our data will also be affected by these pump-induced electric fields since these will be generated while the photoelectrons propagate towards the analyzer as sketched in Fig. \ref{fig:3}(j). In both cases the acceleration of the probe photoelectrons as they propagate in vacuum causes a rigid shift in kinetic energy as well as a broadening in their energy distribution, as observed in Figs. \ref{fig:3}(c) and 3(f), respectively. We note here that a distinction between SPV and vacuum space-charge is difficult from our data. The similar slow dynamics in both regions I and IV suggest that space-charge is the dominating effect since the SPV should diminish within a few picoseconds at positive delay times due to fast recombination of electron-hole pairs in the semiconducting substrate \cite{Yang:2014aa}. Furthermore, SiC has an indirect band gap of $\approx 3$~eV such that electron-hole pair separation by the 1.55~eV pump pulse would require both multiple photon absorption and transfer of momentum by e.g. phonons, which seems unlikely. In the experiments on the observation of the SPV effect in GaAs by S.-L. Yang \emph{et al.}, the effect of space-charge could be ruled out due to the very high repetition rate of the laser (80~MHz) and a low pump fluence (25~$\mu$J/cm$^2$) \cite{Yang:2014aa} in contrast to our experiments.

The change of energy broadening with time, shown in Fig. \ref{fig:3}(f), leads to an estimate of the inherent energy resolution of the experiment of 380~meV. This value is found at long time delays before the energy distributions of the photoemitted electrons are severely affected by the abovementioned pump-induced electric fields. The apparent energy resolution during the intrinsic hot electron signal in graphene corresponds to the plateau reached in Fig. \ref{fig:3}(f) just before the optical excitation, where the pump-induced energy broadening is most severe. Throughout the dynamics in regions II and III it is assumed that this energy broadening remains fixed, and changes in the width of the Fermi edge are fully determined by the variation of the electronic temperature $T_e$.
 
The observed energy shifts and broadenings are a function of applied pump fluence as shown in Figs. \ref{fig:3}(g)-(h). Here, the value for $\Delta E$ is determined from the plateau just before the arrival of the optical excitation. Below $1$~mJ/cm$^2$ the apparent energy resolution is similar to the intrinsic energy resolution we could determine in this experiment, but above this fluence broadening rapidly sets in. Above $4$~mJ/cm$^2$ the broadening is so large that it becomes difficult to discern spectroscopic features. The related chemical potential shift $\Delta\mu$ defined in Fig. \ref{fig:3}(c) shows that the long time delay dynamics in regions I and IV becomes negligible at low fluences, while extensive shifts of up to 1~eV occur above $6$~mJ/cm$^2$. This behavior is consistent with the fact that the multiple photon absorption processes leading to either space-charge or SPV become more efficient with higher fluence. This fluence dependence is highly nonlinear as seen in the data in Figs. \ref{fig:3}(g)-(h). We also observe in the experiments that the energy shift and broadening are very sensitive to both beam spot size and the illuminated area on the sample, which relate to the sharpness of the beam profile and the sample morphology. It is therefore crucial in any experiment to keep the fluence sufficiently low that these effects do not overshadow the intrinsc dynamics in the sample.

\subsection{Double pump excitation of graphene}
\begin{figure} [t!]
\includegraphics[width=0.5\textwidth]{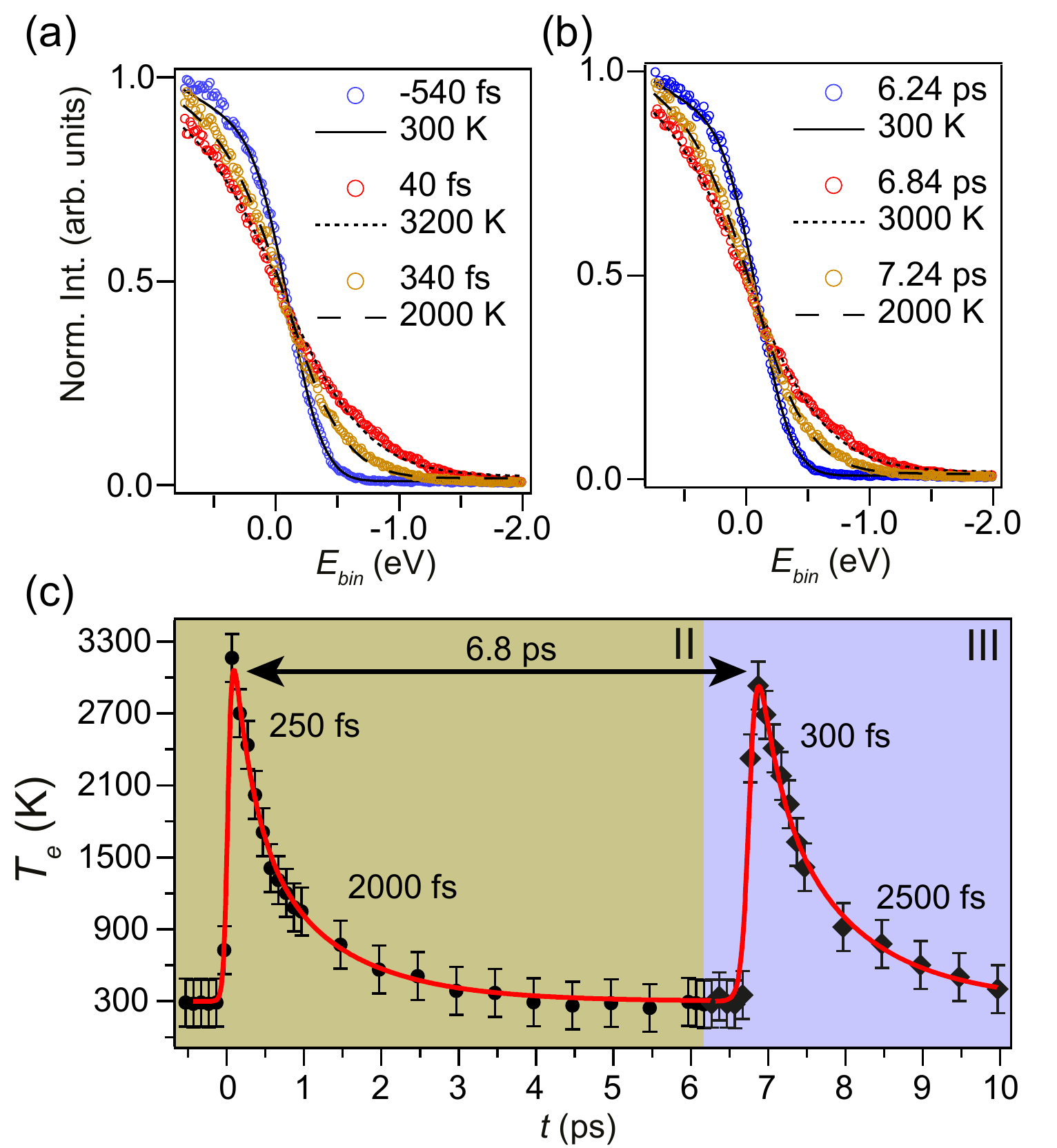}\\
\caption{(a)-(b) MDC-derived energy distribution curves for (a) the first optical excitation and (b) the second excitation. Markers correspond to data at the given time delays and lines are fits to FD functions with the given electronic temperature convoluted with a Gaussian resolution function. (c) Fitted electronic temperature for first (round markers) and second (diamond markers) excitations in regions II and III, respectively. The red lines are exponential function fits with the given time constants for the decay parts. The time difference between the onset of the two excitations is 6.8~ps as marked by the double headed arrow.}
\label{fig:4}
\end{figure}
The electron dynamics in regions II and III reflect the hot electron dynamics in graphene discussed in great detail elsewhere \cite{Johannsen:2013aa,Ulstrup:2015a}. The surprise here is that it occurs twice and with nearly the same intensity. The MDC-derived energy distribution curves in Figs. \ref{fig:4}(a)-(b) have been fitted by FD functions and a Gaussian resolution function with $\Delta E$ set by the apparent resolution immediately before the arrival of the first optical excitation. Since $\mu$ and $T_e$ are free parameters, it is possible to unambiguously extract the time dependence of $T_e$ for the hot electrons in graphene. For each $T_e$ trace in regions II and III, displayed in Fig. \ref{fig:4}(c), the rising edge and the subsequent decay are fitted by a product of exponential functions. The decay part is fitted by two exponential decays, which provide the time constants stated  in Fig. \ref{fig:4}(c). Maximum electronic temperatures of 3200~K and 3000~K are reached with a time difference of 6.8~ps. In both cases a decay with two similar time constants sets in. The fast decay ($\approx 250$~fs) is caused by optical phonons, while the slow decay ($\approx 2000$~fs) is caused by the joint effect of acoustic phonons and impurities \cite{Johannsen:2013aa,Ulstrup:2015a}. 

We are able to confirm that the occurrence of two peaks originates from a back reflection of the pump beam in the substrate by using the measured time delay between the two excitations. The first excitation occurs once the pump reaches the sample, while the second excitation takes place after the pump has traveled back and forth in the SiC crystal. Using a refractive index of 2.6 for SiC at 800~nm, the measured delay of 6.8~ps would correspond to a thickness of $\approx390$~$\mu$m of the sample, which matches the actual thickness of the crystal. By using a sample with a different thickness of the SiC substrate, we have checked that the time delay between the two excitations changed accordingly.
The fact that the maximum electronic temperature is high for both excitations shows that very little energy of the pump is dissipated during its initial interaction with both graphene and SiC. Since two excitations occur, the space-charge and SPV effects are also induced twice. It is difficult to predict how this affects the overall dynamics, and this could play a role for the slightly different decay times extracted for the two excitations in Fig. \ref{fig:4}(c). 

\section{Summary and Perspectives}
Summarizing, we have shown that TR-ARPES spectra of the Dirac cone in monolayer graphene on SiC exhibit strong shifts and broadening in energy. The observed temporal evolution of these perturbations are ascribed to a complex interplay of vacuum space-charge as well as a surface photovoltage in the SiC substrate, induced by the pump beam via multiple photon absorption processes. Additionally, the pump beam is found to be reflected by the back surface of the substrate, giving rise to a duplicate optical excitation signal of hot electrons in graphene.

With the present experimental setup for TR-ARPES such effects are most easily avoided by lowering the fluence of the pump laser into a regime where the SPV and space-charge effects are negligible compared to the intrinsic electron dynamics of the sample. A further exploration of the SPV effect, e.g. to disentangle its contribution to the dynamics on the long picosecond time scales from the photoelectron space-charge cloud, would require a substrate with a different band gap. Alternatively, one could use graphene on a metal substrate to avoid SPV altogether, but this reduces the pump-probe signal due to screening of the pump laser field by the metal \cite{Ulstrup:2015a}. These issues with pump (and probe) induced external electric fields that interfere with the photoemission process are expected to become less of a problem with future TR-ARPES setups utilizing laser sources with much higher repetition rates since the energy per pulse will be significantly reduced.

\section{Acknowledgements}
We gratefully acknowledge financial support from the VILLUM foundation, The Danish Council for Independent Research / Technology and Production Sciences, the Swiss National Science Foundation (NSF), EPSRC, The Royal Society and the Italian Ministry of University and Research (Grants No. FIRBRBAP045JF2 and No. FIRB-RBAP06AWK3). Access to the Artemis Facility was funded by STFC. Work in Chemnitz was supported by the European Union through the project ConceptGraphene, and by the German Research Foundation in the framework of the SPP 1459 Graphene.

%

\begin{thebibliography}{44}
\expandafter\ifx\csname natexlab\endcsname\relax\def\natexlab#1{#1}\fi
\expandafter\ifx\csname bibnamefont\endcsname\relax
  \def\bibnamefont#1{#1}\fi
\expandafter\ifx\csname bibfnamefont\endcsname\relax
  \def\bibfnamefont#1{#1}\fi
\expandafter\ifx\csname citenamefont\endcsname\relax
  \def\citenamefont#1{#1}\fi
\expandafter\ifx\csname url\endcsname\relax
  \def\url#1{\texttt{#1}}\fi
\expandafter\ifx\csname urlprefix\endcsname\relax\def\urlprefix{URL }\fi
\providecommand{\bibinfo}[2]{#2}
\providecommand{\eprint}[2][]{\url{#2}}

\bibitem[{\citenamefont{Perfetti et~al.}(2007)\citenamefont{Perfetti, Loukakos,
  Lisowski, Bovensiepen, Eisaki, and Wolf}}]{Perfetti:2007}
\bibinfo{author}{\bibfnamefont{L.}~\bibnamefont{Perfetti}},
  \bibinfo{author}{\bibfnamefont{P.~A.} \bibnamefont{Loukakos}},
  \bibinfo{author}{\bibfnamefont{M.}~\bibnamefont{Lisowski}},
  \bibinfo{author}{\bibfnamefont{U.}~\bibnamefont{Bovensiepen}},
  \bibinfo{author}{\bibfnamefont{H.}~\bibnamefont{Eisaki}}, \bibnamefont{and}
  \bibinfo{author}{\bibfnamefont{M.}~\bibnamefont{Wolf}},
  \bibinfo{journal}{Phys. Rev. Lett.} \textbf{\bibinfo{volume}{99}},
  \bibinfo{pages}{197001} (\bibinfo{year}{2007}),
  \urlprefix\url{http://link.aps.org/doi/10.1103/PhysRevLett.99.197001}.

\bibitem[{\citenamefont{Schmitt et~al.}(2008)\citenamefont{Schmitt, Kirchmann,
  Bovensiepen, Moore, Rettig, Krenz, Chu, Ru, Perfetti, Lu
  et~al.}}]{Schmitt:2008}
\bibinfo{author}{\bibfnamefont{F.}~\bibnamefont{Schmitt}},
  \bibinfo{author}{\bibfnamefont{P.~S.} \bibnamefont{Kirchmann}},
  \bibinfo{author}{\bibfnamefont{U.}~\bibnamefont{Bovensiepen}},
  \bibinfo{author}{\bibfnamefont{R.~G.} \bibnamefont{Moore}},
  \bibinfo{author}{\bibfnamefont{L.}~\bibnamefont{Rettig}},
  \bibinfo{author}{\bibfnamefont{M.}~\bibnamefont{Krenz}},
  \bibinfo{author}{\bibfnamefont{J.-H.} \bibnamefont{Chu}},
  \bibinfo{author}{\bibfnamefont{N.}~\bibnamefont{Ru}},
  \bibinfo{author}{\bibfnamefont{L.}~\bibnamefont{Perfetti}},
  \bibinfo{author}{\bibfnamefont{D.~H.} \bibnamefont{Lu}},
  \bibnamefont{et~al.}, \bibinfo{journal}{Science}
  \textbf{\bibinfo{volume}{321}}, \bibinfo{pages}{1649} (\bibinfo{year}{2008}),
  \eprint{http://www.sciencemag.org/content/321/5896/1649.full.pdf},
  \urlprefix\url{http://www.sciencemag.org/content/321/5896/1649.abstract}.

\bibitem[{\citenamefont{Petersen et~al.}(2011)\citenamefont{Petersen, Kaiser,
  Dean, Simoncig, Liu, Cavalieri, Cacho, Turcu, Springate, Frassetto
  et~al.}}]{Petersen:2011}
\bibinfo{author}{\bibfnamefont{J.~C.} \bibnamefont{Petersen}},
  \bibinfo{author}{\bibfnamefont{S.}~\bibnamefont{Kaiser}},
  \bibinfo{author}{\bibfnamefont{N.}~\bibnamefont{Dean}},
  \bibinfo{author}{\bibfnamefont{A.}~\bibnamefont{Simoncig}},
  \bibinfo{author}{\bibfnamefont{H.~Y.} \bibnamefont{Liu}},
  \bibinfo{author}{\bibfnamefont{A.~L.} \bibnamefont{Cavalieri}},
  \bibinfo{author}{\bibfnamefont{C.}~\bibnamefont{Cacho}},
  \bibinfo{author}{\bibfnamefont{I.~C.~E.} \bibnamefont{Turcu}},
  \bibinfo{author}{\bibfnamefont{E.}~\bibnamefont{Springate}},
  \bibinfo{author}{\bibfnamefont{F.}~\bibnamefont{Frassetto}},
  \bibnamefont{et~al.}, \bibinfo{journal}{Phys. Rev. Lett.}
  \textbf{\bibinfo{volume}{107}}, \bibinfo{pages}{177402}
  (\bibinfo{year}{2011}),
  \urlprefix\url{http://link.aps.org/doi/10.1103/PhysRevLett.107.177402}.

\bibitem[{\citenamefont{Rohwer et~al.}(2011)\citenamefont{Rohwer, Hellmann,
  Wiesenmayer, Sohrt, Stange, Slomski, Carr, Liu, Avila, Kallane
  et~al.}}]{Rohwer:2011}
\bibinfo{author}{\bibfnamefont{T.}~\bibnamefont{Rohwer}},
  \bibinfo{author}{\bibfnamefont{S.}~\bibnamefont{Hellmann}},
  \bibinfo{author}{\bibfnamefont{M.}~\bibnamefont{Wiesenmayer}},
  \bibinfo{author}{\bibfnamefont{C.}~\bibnamefont{Sohrt}},
  \bibinfo{author}{\bibfnamefont{A.}~\bibnamefont{Stange}},
  \bibinfo{author}{\bibfnamefont{B.}~\bibnamefont{Slomski}},
  \bibinfo{author}{\bibfnamefont{A.}~\bibnamefont{Carr}},
  \bibinfo{author}{\bibfnamefont{Y.}~\bibnamefont{Liu}},
  \bibinfo{author}{\bibfnamefont{L.~M.} \bibnamefont{Avila}},
  \bibinfo{author}{\bibfnamefont{M.}~\bibnamefont{Kallane}},
  \bibnamefont{et~al.}, \bibinfo{journal}{Nature}
  \textbf{\bibinfo{volume}{471}}, \bibinfo{pages}{490} (\bibinfo{year}{2011}),
  \urlprefix\url{http://dx.doi.org/10.1038/nature09829}.

\bibitem[{\citenamefont{Hellmann et~al.}(2010)\citenamefont{Hellmann, Beye,
  Sohrt, Rohwer, Sorgenfrei, Redlin, Kall\"ane, Marczynski-B\"uhlow, Hennies,
  Bauer et~al.}}]{Hellmann:2010}
\bibinfo{author}{\bibfnamefont{S.}~\bibnamefont{Hellmann}},
  \bibinfo{author}{\bibfnamefont{M.}~\bibnamefont{Beye}},
  \bibinfo{author}{\bibfnamefont{C.}~\bibnamefont{Sohrt}},
  \bibinfo{author}{\bibfnamefont{T.}~\bibnamefont{Rohwer}},
  \bibinfo{author}{\bibfnamefont{F.}~\bibnamefont{Sorgenfrei}},
  \bibinfo{author}{\bibfnamefont{H.}~\bibnamefont{Redlin}},
  \bibinfo{author}{\bibfnamefont{M.}~\bibnamefont{Kall\"ane}},
  \bibinfo{author}{\bibfnamefont{M.}~\bibnamefont{Marczynski-B\"uhlow}},
  \bibinfo{author}{\bibfnamefont{F.}~\bibnamefont{Hennies}},
  \bibinfo{author}{\bibfnamefont{M.}~\bibnamefont{Bauer}},
  \bibnamefont{et~al.}, \bibinfo{journal}{Phys. Rev. Lett.}
  \textbf{\bibinfo{volume}{105}}, \bibinfo{pages}{187401}
  (\bibinfo{year}{2010}),
  \urlprefix\url{http://link.aps.org/doi/10.1103/PhysRevLett.105.187401}.

\bibitem[{\citenamefont{Cort\'es et~al.}(2011)\citenamefont{Cort\'es, Rettig,
  Yoshida, Eisaki, Wolf, and Bovensiepen}}]{Cortes:2011}
\bibinfo{author}{\bibfnamefont{R.}~\bibnamefont{Cort\'es}},
  \bibinfo{author}{\bibfnamefont{L.}~\bibnamefont{Rettig}},
  \bibinfo{author}{\bibfnamefont{Y.}~\bibnamefont{Yoshida}},
  \bibinfo{author}{\bibfnamefont{H.}~\bibnamefont{Eisaki}},
  \bibinfo{author}{\bibfnamefont{M.}~\bibnamefont{Wolf}}, \bibnamefont{and}
  \bibinfo{author}{\bibfnamefont{U.}~\bibnamefont{Bovensiepen}},
  \bibinfo{journal}{Phys. Rev. Lett.} \textbf{\bibinfo{volume}{107}},
  \bibinfo{pages}{097002} (\bibinfo{year}{2011}),
  \urlprefix\url{http://link.aps.org/doi/10.1103/PhysRevLett.107.097002}.

\bibitem[{\citenamefont{Graf et~al.}(2011)\citenamefont{Graf, Jozwiak,
  Smallwood, Eisaki, Kaindl, Lee, and Lanzara}}]{Graf:2011aa}
\bibinfo{author}{\bibfnamefont{J.}~\bibnamefont{Graf}},
  \bibinfo{author}{\bibfnamefont{C.}~\bibnamefont{Jozwiak}},
  \bibinfo{author}{\bibfnamefont{C.~L.} \bibnamefont{Smallwood}},
  \bibinfo{author}{\bibfnamefont{H.}~\bibnamefont{Eisaki}},
  \bibinfo{author}{\bibfnamefont{R.~A.} \bibnamefont{Kaindl}},
  \bibinfo{author}{\bibfnamefont{D.-H.} \bibnamefont{Lee}}, \bibnamefont{and}
  \bibinfo{author}{\bibfnamefont{A.}~\bibnamefont{Lanzara}},
  \bibinfo{journal}{Nat Phys} \textbf{\bibinfo{volume}{7}},
  \bibinfo{pages}{805} (\bibinfo{year}{2011}),
  \urlprefix\url{http://dx.doi.org/10.1038/nphys2027}.

\bibitem[{\citenamefont{Smallwood et~al.}(2012)\citenamefont{Smallwood, Hinton,
  Jozwiak, Zhang, Koralek, Eisaki, Lee, Orenstein, and
  Lanzara}}]{Smallwood:2012}
\bibinfo{author}{\bibfnamefont{C.~L.} \bibnamefont{Smallwood}},
  \bibinfo{author}{\bibfnamefont{J.~P.} \bibnamefont{Hinton}},
  \bibinfo{author}{\bibfnamefont{C.}~\bibnamefont{Jozwiak}},
  \bibinfo{author}{\bibfnamefont{W.}~\bibnamefont{Zhang}},
  \bibinfo{author}{\bibfnamefont{J.~D.} \bibnamefont{Koralek}},
  \bibinfo{author}{\bibfnamefont{H.}~\bibnamefont{Eisaki}},
  \bibinfo{author}{\bibfnamefont{D.-H.} \bibnamefont{Lee}},
  \bibinfo{author}{\bibfnamefont{J.}~\bibnamefont{Orenstein}},
  \bibnamefont{and} \bibinfo{author}{\bibfnamefont{A.}~\bibnamefont{Lanzara}},
  \bibinfo{journal}{Science} \textbf{\bibinfo{volume}{336}},
  \bibinfo{pages}{1137} (\bibinfo{year}{2012}),
  \eprint{http://www.sciencemag.org/content/336/6085/1137.full.pdf},
  \urlprefix\url{http://www.sciencemag.org/content/336/6085/1137.abstract}.

\bibitem[{\citenamefont{Johannsen
  et~al.}(2013{\natexlab{a}})\citenamefont{Johannsen, Ulstrup, Cilento,
  Crepaldi, Zacchigna, Cacho, Turcu, Springate, Fromm, Raidel
  et~al.}}]{Johannsen:2013aa}
\bibinfo{author}{\bibfnamefont{J.~C.} \bibnamefont{Johannsen}},
  \bibinfo{author}{\bibfnamefont{S.}~\bibnamefont{Ulstrup}},
  \bibinfo{author}{\bibfnamefont{F.}~\bibnamefont{Cilento}},
  \bibinfo{author}{\bibfnamefont{A.}~\bibnamefont{Crepaldi}},
  \bibinfo{author}{\bibfnamefont{M.}~\bibnamefont{Zacchigna}},
  \bibinfo{author}{\bibfnamefont{C.}~\bibnamefont{Cacho}},
  \bibinfo{author}{\bibfnamefont{I.~C.~E.} \bibnamefont{Turcu}},
  \bibinfo{author}{\bibfnamefont{E.}~\bibnamefont{Springate}},
  \bibinfo{author}{\bibfnamefont{F.}~\bibnamefont{Fromm}},
  \bibinfo{author}{\bibfnamefont{C.}~\bibnamefont{Raidel}},
  \bibnamefont{et~al.}, \bibinfo{journal}{Phys. Rev. Lett.}
  \textbf{\bibinfo{volume}{111}}, \bibinfo{pages}{027403}
  (\bibinfo{year}{2013}{\natexlab{a}}),
  \urlprefix\url{http://link.aps.org/doi/10.1103/PhysRevLett.111.027403}.

\bibitem[{\citenamefont{Ulstrup
  et~al.}(2014{\natexlab{a}})\citenamefont{Ulstrup, Johannsen, Cilento, Miwa,
  Crepaldi, Zacchigna, Cacho, Chapman, Springate, Mammadov
  et~al.}}]{Ulstrup:2014b}
\bibinfo{author}{\bibfnamefont{S.}~\bibnamefont{Ulstrup}},
  \bibinfo{author}{\bibfnamefont{J.~C.} \bibnamefont{Johannsen}},
  \bibinfo{author}{\bibfnamefont{F.}~\bibnamefont{Cilento}},
  \bibinfo{author}{\bibfnamefont{J.~A.} \bibnamefont{Miwa}},
  \bibinfo{author}{\bibfnamefont{A.}~\bibnamefont{Crepaldi}},
  \bibinfo{author}{\bibfnamefont{M.}~\bibnamefont{Zacchigna}},
  \bibinfo{author}{\bibfnamefont{C.}~\bibnamefont{Cacho}},
  \bibinfo{author}{\bibfnamefont{R.}~\bibnamefont{Chapman}},
  \bibinfo{author}{\bibfnamefont{E.}~\bibnamefont{Springate}},
  \bibinfo{author}{\bibfnamefont{S.}~\bibnamefont{Mammadov}},
  \bibnamefont{et~al.}, \bibinfo{journal}{Phys. Rev. Lett.}
  \textbf{\bibinfo{volume}{112}}, \bibinfo{pages}{257401}
  (\bibinfo{year}{2014}{\natexlab{a}}),
  \urlprefix\url{http://link.aps.org/doi/10.1103/PhysRevLett.112.257401}.

\bibitem[{\citenamefont{Gierz et~al.}(2013)\citenamefont{Gierz, Petersen,
  Mitrano, Cacho, Turcu, Springate, St{\"o}hr, K{\"o}hler, Starke, and
  Cavalleri}}]{Gierz:2013aa}
\bibinfo{author}{\bibfnamefont{I.}~\bibnamefont{Gierz}},
  \bibinfo{author}{\bibfnamefont{J.~C.} \bibnamefont{Petersen}},
  \bibinfo{author}{\bibfnamefont{M.}~\bibnamefont{Mitrano}},
  \bibinfo{author}{\bibfnamefont{C.}~\bibnamefont{Cacho}},
  \bibinfo{author}{\bibfnamefont{I.~C.~E.} \bibnamefont{Turcu}},
  \bibinfo{author}{\bibfnamefont{E.}~\bibnamefont{Springate}},
  \bibinfo{author}{\bibfnamefont{A.}~\bibnamefont{St{\"o}hr}},
  \bibinfo{author}{\bibfnamefont{A.}~\bibnamefont{K{\"o}hler}},
  \bibinfo{author}{\bibfnamefont{U.}~\bibnamefont{Starke}}, \bibnamefont{and}
  \bibinfo{author}{\bibfnamefont{A.}~\bibnamefont{Cavalleri}},
  \bibinfo{journal}{Nat Mater} \textbf{\bibinfo{volume}{12}},
  \bibinfo{pages}{1119} (\bibinfo{year}{2013}),
  \urlprefix\url{http://dx.doi.org/10.1038/nmat3757}.

\bibitem[{\citenamefont{Johannsen et~al.}(2015)\citenamefont{Johannsen,
  Ulstrup, Crepaldi, Cilento, Zacchigna, Miwa, Cacho, Chapman, Springate, Fromm
  et~al.}}]{Johannsen:2015}
\bibinfo{author}{\bibfnamefont{J.~C.} \bibnamefont{Johannsen}},
  \bibinfo{author}{\bibfnamefont{S.}~\bibnamefont{Ulstrup}},
  \bibinfo{author}{\bibfnamefont{A.}~\bibnamefont{Crepaldi}},
  \bibinfo{author}{\bibfnamefont{F.}~\bibnamefont{Cilento}},
  \bibinfo{author}{\bibfnamefont{M.}~\bibnamefont{Zacchigna}},
  \bibinfo{author}{\bibfnamefont{J.~A.} \bibnamefont{Miwa}},
  \bibinfo{author}{\bibfnamefont{C.}~\bibnamefont{Cacho}},
  \bibinfo{author}{\bibfnamefont{R.~T.} \bibnamefont{Chapman}},
  \bibinfo{author}{\bibfnamefont{E.}~\bibnamefont{Springate}},
  \bibinfo{author}{\bibfnamefont{F.}~\bibnamefont{Fromm}},
  \bibnamefont{et~al.}, \bibinfo{journal}{Nano Letters}
  \textbf{\bibinfo{volume}{15}}, \bibinfo{pages}{326} (\bibinfo{year}{2015}),
  \bibinfo{note}{pMID: 25458168}, \eprint{http://dx.doi.org/10.1021/nl503614v},
  \urlprefix\url{http://dx.doi.org/10.1021/nl503614v}.

\bibitem[{\citenamefont{Ulstrup et~al.}(2015)\citenamefont{Ulstrup, Johannsen,
  Crepaldi, Cilento, Zacchigna, Cacho, Chapman, Springate, Fromm, Raidel
  et~al.}}]{Ulstrup:2015a}
\bibinfo{author}{\bibfnamefont{S.}~\bibnamefont{Ulstrup}},
  \bibinfo{author}{\bibfnamefont{J.~C.} \bibnamefont{Johannsen}},
  \bibinfo{author}{\bibfnamefont{A.}~\bibnamefont{Crepaldi}},
  \bibinfo{author}{\bibfnamefont{F.}~\bibnamefont{Cilento}},
  \bibinfo{author}{\bibfnamefont{M.}~\bibnamefont{Zacchigna}},
  \bibinfo{author}{\bibfnamefont{C.}~\bibnamefont{Cacho}},
  \bibinfo{author}{\bibfnamefont{R.}~\bibnamefont{Chapman}},
  \bibinfo{author}{\bibfnamefont{E.}~\bibnamefont{Springate}},
  \bibinfo{author}{\bibfnamefont{F.}~\bibnamefont{Fromm}},
  \bibinfo{author}{\bibfnamefont{C.}~\bibnamefont{Raidel}},
  \bibnamefont{et~al.}, \bibinfo{journal}{J. Phys. Cond. Matter}
  (\bibinfo{year}{2015}).

\bibitem[{\citenamefont{Crepaldi et~al.}(2012)\citenamefont{Crepaldi, Ressel,
  Cilento, Zacchigna, Grazioli, Berger, Bugnon, Kern, Grioni, and
  Parmigiani}}]{Crepaldi:2012}
\bibinfo{author}{\bibfnamefont{A.}~\bibnamefont{Crepaldi}},
  \bibinfo{author}{\bibfnamefont{B.}~\bibnamefont{Ressel}},
  \bibinfo{author}{\bibfnamefont{F.}~\bibnamefont{Cilento}},
  \bibinfo{author}{\bibfnamefont{M.}~\bibnamefont{Zacchigna}},
  \bibinfo{author}{\bibfnamefont{C.}~\bibnamefont{Grazioli}},
  \bibinfo{author}{\bibfnamefont{H.}~\bibnamefont{Berger}},
  \bibinfo{author}{\bibfnamefont{P.}~\bibnamefont{Bugnon}},
  \bibinfo{author}{\bibfnamefont{K.}~\bibnamefont{Kern}},
  \bibinfo{author}{\bibfnamefont{M.}~\bibnamefont{Grioni}}, \bibnamefont{and}
  \bibinfo{author}{\bibfnamefont{F.}~\bibnamefont{Parmigiani}},
  \bibinfo{journal}{Phys. Rev. B} \textbf{\bibinfo{volume}{86}},
  \bibinfo{pages}{205133} (\bibinfo{year}{2012}),
  \urlprefix\url{http://link.aps.org/doi/10.1103/PhysRevB.86.205133}.

\bibitem[{\citenamefont{Crepaldi et~al.}(2013)\citenamefont{Crepaldi, Cilento,
  Ressel, Cacho, Johannsen, Zacchigna, Berger, Bugnon, Grazioli, Turcu
  et~al.}}]{crepaldi2013}
\bibinfo{author}{\bibfnamefont{A.}~\bibnamefont{Crepaldi}},
  \bibinfo{author}{\bibfnamefont{F.}~\bibnamefont{Cilento}},
  \bibinfo{author}{\bibfnamefont{B.}~\bibnamefont{Ressel}},
  \bibinfo{author}{\bibfnamefont{C.}~\bibnamefont{Cacho}},
  \bibinfo{author}{\bibfnamefont{J.~C.} \bibnamefont{Johannsen}},
  \bibinfo{author}{\bibfnamefont{M.}~\bibnamefont{Zacchigna}},
  \bibinfo{author}{\bibfnamefont{H.}~\bibnamefont{Berger}},
  \bibinfo{author}{\bibfnamefont{P.}~\bibnamefont{Bugnon}},
  \bibinfo{author}{\bibfnamefont{C.}~\bibnamefont{Grazioli}},
  \bibinfo{author}{\bibfnamefont{I.~C.~E.} \bibnamefont{Turcu}},
  \bibnamefont{et~al.}, \bibinfo{journal}{Physical Review B}
  \textbf{\bibinfo{volume}{88}}, \bibinfo{pages}{121404(R)}
  (\bibinfo{year}{2013}),
  \urlprefix\url{http://dx.doi.org/10.1103/PhysRevB.88.121404}.

\bibitem[{\citenamefont{Sobota et~al.}(2012)\citenamefont{Sobota, Yang,
  Analytis, Chen, Fisher, Kirchmann, and Shen}}]{Sobota:2012}
\bibinfo{author}{\bibfnamefont{J.~A.} \bibnamefont{Sobota}},
  \bibinfo{author}{\bibfnamefont{S.}~\bibnamefont{Yang}},
  \bibinfo{author}{\bibfnamefont{J.~G.} \bibnamefont{Analytis}},
  \bibinfo{author}{\bibfnamefont{Y.~L.} \bibnamefont{Chen}},
  \bibinfo{author}{\bibfnamefont{I.~R.} \bibnamefont{Fisher}},
  \bibinfo{author}{\bibfnamefont{P.~S.} \bibnamefont{Kirchmann}},
  \bibnamefont{and} \bibinfo{author}{\bibfnamefont{Z.-X.} \bibnamefont{Shen}},
  \bibinfo{journal}{Phys. Rev. Lett.} \textbf{\bibinfo{volume}{108}},
  \bibinfo{pages}{117403} (\bibinfo{year}{2012}),
  \urlprefix\url{http://link.aps.org/doi/10.1103/PhysRevLett.108.117403}.

\bibitem[{\citenamefont{Wang et~al.}(2012)\citenamefont{Wang, Hsieh, Sie,
  Steinberg, Gardner, Lee, Jarillo-Herrero, and Gedik}}]{Wang:2012d}
\bibinfo{author}{\bibfnamefont{Y.~H.} \bibnamefont{Wang}},
  \bibinfo{author}{\bibfnamefont{D.}~\bibnamefont{Hsieh}},
  \bibinfo{author}{\bibfnamefont{E.~J.} \bibnamefont{Sie}},
  \bibinfo{author}{\bibfnamefont{H.}~\bibnamefont{Steinberg}},
  \bibinfo{author}{\bibfnamefont{D.~R.} \bibnamefont{Gardner}},
  \bibinfo{author}{\bibfnamefont{Y.~S.} \bibnamefont{Lee}},
  \bibinfo{author}{\bibfnamefont{P.}~\bibnamefont{Jarillo-Herrero}},
  \bibnamefont{and} \bibinfo{author}{\bibfnamefont{N.}~\bibnamefont{Gedik}},
  \bibinfo{journal}{Phys. Rev. Lett.} \textbf{\bibinfo{volume}{109}},
  \bibinfo{pages}{127401} (\bibinfo{year}{2012}),
  \urlprefix\url{http://link.aps.org/doi/10.1103/PhysRevLett.109.127401}.

\bibitem[{\citenamefont{Wang et~al.}(2013)\citenamefont{Wang, Steinberg,
  Jarillo-Herrero, and Gedik}}]{Wang:2013c}
\bibinfo{author}{\bibfnamefont{Y.~H.} \bibnamefont{Wang}},
  \bibinfo{author}{\bibfnamefont{H.}~\bibnamefont{Steinberg}},
  \bibinfo{author}{\bibfnamefont{P.}~\bibnamefont{Jarillo-Herrero}},
  \bibnamefont{and} \bibinfo{author}{\bibfnamefont{N.}~\bibnamefont{Gedik}},
  \bibinfo{journal}{Science} \textbf{\bibinfo{volume}{342}},
  \bibinfo{pages}{453} (\bibinfo{year}{2013}),
  \eprint{http://www.sciencemag.org/content/342/6157/453.full.pdf},
  \urlprefix\url{http://www.sciencemag.org/content/342/6157/453.abstract}.

\bibitem[{\citenamefont{Hajlaoui et~al.}(2014)\citenamefont{Hajlaoui,
  Papalazarou, Mauchain, Perfetti, Taleb-Ibrahimi, Navarin, Monteverde,
  Auban-Senzier, Pasquier, Moisan et~al.}}]{Hajlaoui:2014aa}
\bibinfo{author}{\bibfnamefont{M.}~\bibnamefont{Hajlaoui}},
  \bibinfo{author}{\bibfnamefont{E.}~\bibnamefont{Papalazarou}},
  \bibinfo{author}{\bibfnamefont{J.}~\bibnamefont{Mauchain}},
  \bibinfo{author}{\bibfnamefont{L.}~\bibnamefont{Perfetti}},
  \bibinfo{author}{\bibfnamefont{A.}~\bibnamefont{Taleb-Ibrahimi}},
  \bibinfo{author}{\bibfnamefont{F.}~\bibnamefont{Navarin}},
  \bibinfo{author}{\bibfnamefont{M.}~\bibnamefont{Monteverde}},
  \bibinfo{author}{\bibfnamefont{P.}~\bibnamefont{Auban-Senzier}},
  \bibinfo{author}{\bibfnamefont{C.~R.} \bibnamefont{Pasquier}},
  \bibinfo{author}{\bibfnamefont{N.}~\bibnamefont{Moisan}},
  \bibnamefont{et~al.}, \bibinfo{journal}{Nat Commun}
  \textbf{\bibinfo{volume}{5}} (\bibinfo{year}{2014}),
  \urlprefix\url{http://dx.doi.org/10.1038/ncomms4003}.

\bibitem[{\citenamefont{Rana}(2007)}]{Rana:2007}
\bibinfo{author}{\bibfnamefont{F.}~\bibnamefont{Rana}}, \bibinfo{journal}{Phys.
  Rev. B} \textbf{\bibinfo{volume}{76}}, \bibinfo{pages}{155431}
  (\bibinfo{year}{2007}),
  \urlprefix\url{http://link.aps.org/doi/10.1103/PhysRevB.76.155431}.

\bibitem[{\citenamefont{Winzer et~al.}(2010)\citenamefont{Winzer, Knorr, and
  Malic}}]{Winzer:2010}
\bibinfo{author}{\bibfnamefont{T.}~\bibnamefont{Winzer}},
  \bibinfo{author}{\bibfnamefont{A.}~\bibnamefont{Knorr}}, \bibnamefont{and}
  \bibinfo{author}{\bibfnamefont{E.}~\bibnamefont{Malic}},
  \bibinfo{journal}{Nano Letters} \textbf{\bibinfo{volume}{10}},
  \bibinfo{pages}{4839} (\bibinfo{year}{2010}),
  \eprint{http://pubs.acs.org/doi/pdf/10.1021/nl1024485},
  \urlprefix\url{http://pubs.acs.org/doi/abs/10.1021/nl1024485}.

\bibitem[{\citenamefont{Song et~al.}(2013)\citenamefont{Song, Tielrooij,
  Koppens, and Levitov}}]{Song:2013}
\bibinfo{author}{\bibfnamefont{J.~C.~W.} \bibnamefont{Song}},
  \bibinfo{author}{\bibfnamefont{K.~J.} \bibnamefont{Tielrooij}},
  \bibinfo{author}{\bibfnamefont{F.~H.~L.} \bibnamefont{Koppens}},
  \bibnamefont{and} \bibinfo{author}{\bibfnamefont{L.~S.}
  \bibnamefont{Levitov}}, \bibinfo{journal}{Phys. Rev. B}
  \textbf{\bibinfo{volume}{87}}, \bibinfo{pages}{155429}
  (\bibinfo{year}{2013}),
  \urlprefix\url{http://link.aps.org/doi/10.1103/PhysRevB.87.155429}.

\bibitem[{\citenamefont{Brida et~al.}(2013)\citenamefont{Brida, Tomadin,
  Manzoni, Kim, Lombardo, Milana, Nair, Novoselov, Ferrari, Cerullo
  et~al.}}]{Brida:2013aa}
\bibinfo{author}{\bibfnamefont{D.}~\bibnamefont{Brida}},
  \bibinfo{author}{\bibfnamefont{A.}~\bibnamefont{Tomadin}},
  \bibinfo{author}{\bibfnamefont{C.}~\bibnamefont{Manzoni}},
  \bibinfo{author}{\bibfnamefont{Y.~J.} \bibnamefont{Kim}},
  \bibinfo{author}{\bibfnamefont{A.}~\bibnamefont{Lombardo}},
  \bibinfo{author}{\bibfnamefont{S.}~\bibnamefont{Milana}},
  \bibinfo{author}{\bibfnamefont{R.~R.} \bibnamefont{Nair}},
  \bibinfo{author}{\bibfnamefont{K.~S.} \bibnamefont{Novoselov}},
  \bibinfo{author}{\bibfnamefont{A.~C.} \bibnamefont{Ferrari}},
  \bibinfo{author}{\bibfnamefont{G.}~\bibnamefont{Cerullo}},
  \bibnamefont{et~al.}, \bibinfo{journal}{Nat Commun}
  \textbf{\bibinfo{volume}{4}}, \bibinfo{pages}{1987} (\bibinfo{year}{2013}),
  \urlprefix\url{http://dx.doi.org/10.1038/ncomms2987}.

\bibitem[{\citenamefont{Tielrooij et~al.}(2013)\citenamefont{Tielrooij, Song,
  Jensen, Centeno, Pesquera, Elorza, Bonn, Levitov, and
  Koppens}}]{Tielrooij:2012}
\bibinfo{author}{\bibfnamefont{K.~J.} \bibnamefont{Tielrooij}},
  \bibinfo{author}{\bibfnamefont{J.~C.~W.} \bibnamefont{Song}},
  \bibinfo{author}{\bibfnamefont{S.~A.} \bibnamefont{Jensen}},
  \bibinfo{author}{\bibfnamefont{A.}~\bibnamefont{Centeno}},
  \bibinfo{author}{\bibfnamefont{A.}~\bibnamefont{Pesquera}},
  \bibinfo{author}{\bibfnamefont{A.~Z.} \bibnamefont{Elorza}},
  \bibinfo{author}{\bibfnamefont{M.}~\bibnamefont{Bonn}},
  \bibinfo{author}{\bibfnamefont{L.~S.} \bibnamefont{Levitov}},
  \bibnamefont{and} \bibinfo{author}{\bibfnamefont{F.~H.~L.}
  \bibnamefont{Koppens}}, \bibinfo{journal}{Nature Physics}
  \textbf{\bibinfo{volume}{9}}, \bibinfo{pages}{248} (\bibinfo{year}{2013}).

\bibitem[{\citenamefont{Pl{\"o}tzing et~al.}(2014)\citenamefont{Pl{\"o}tzing,
  Winzer, Malic, Neumaier, Knorr, and Kurz}}]{Ploetzing:2014}
\bibinfo{author}{\bibfnamefont{T.}~\bibnamefont{Pl{\"o}tzing}},
  \bibinfo{author}{\bibfnamefont{T.}~\bibnamefont{Winzer}},
  \bibinfo{author}{\bibfnamefont{E.}~\bibnamefont{Malic}},
  \bibinfo{author}{\bibfnamefont{D.}~\bibnamefont{Neumaier}},
  \bibinfo{author}{\bibfnamefont{A.}~\bibnamefont{Knorr}}, \bibnamefont{and}
  \bibinfo{author}{\bibfnamefont{H.}~\bibnamefont{Kurz}},
  \bibinfo{journal}{Nano Letters} \textbf{\bibinfo{volume}{14}},
  \bibinfo{pages}{5371} (\bibinfo{year}{2014}),
  \eprint{http://dx.doi.org/10.1021/nl502114w},
  \urlprefix\url{http://dx.doi.org/10.1021/nl502114w}.

\bibitem[{\citenamefont{Song et~al.}(2012)\citenamefont{Song, Reizer, and
  Levitov}}]{Song:2012c}
\bibinfo{author}{\bibfnamefont{J.~C.~W.} \bibnamefont{Song}},
  \bibinfo{author}{\bibfnamefont{M.~Y.} \bibnamefont{Reizer}},
  \bibnamefont{and} \bibinfo{author}{\bibfnamefont{L.~S.}
  \bibnamefont{Levitov}}, \bibinfo{journal}{Phys. Rev. Lett.}
  \textbf{\bibinfo{volume}{109}}, \bibinfo{pages}{106602}
  (\bibinfo{year}{2012}),
  \urlprefix\url{http://link.aps.org/doi/10.1103/PhysRevLett.109.106602}.

\bibitem[{\citenamefont{Bistritzer and MacDonald}(2009)}]{Bistritzer:2009}
\bibinfo{author}{\bibfnamefont{R.}~\bibnamefont{Bistritzer}} \bibnamefont{and}
  \bibinfo{author}{\bibfnamefont{A.~H.} \bibnamefont{MacDonald}},
  \bibinfo{journal}{Phys. Rev. Lett.} \textbf{\bibinfo{volume}{102}},
  \bibinfo{pages}{206410} (\bibinfo{year}{2009}),
  \urlprefix\url{http://link.aps.org/doi/10.1103/PhysRevLett.102.206410}.

\bibitem[{\citenamefont{Butscher et~al.}(2007)\citenamefont{Butscher, Milde,
  Hirtschulz, Malic, and Knorr}}]{Butscher:2007}
\bibinfo{author}{\bibfnamefont{S.}~\bibnamefont{Butscher}},
  \bibinfo{author}{\bibfnamefont{F.}~\bibnamefont{Milde}},
  \bibinfo{author}{\bibfnamefont{M.}~\bibnamefont{Hirtschulz}},
  \bibinfo{author}{\bibfnamefont{E.}~\bibnamefont{Malic}}, \bibnamefont{and}
  \bibinfo{author}{\bibfnamefont{A.}~\bibnamefont{Knorr}},
  \bibinfo{journal}{Applied Physics Letters} \textbf{\bibinfo{volume}{91}},
  \bibinfo{eid}{203103} (pages~\bibinfo{numpages}{3}) (\bibinfo{year}{2007}),
  \urlprefix\url{http://link.aip.org/link/?APL/91/203103/1}.

\bibitem[{\citenamefont{Zhou et~al.}(2005)\citenamefont{Zhou, Wannberg, Yang,
  Brouet, Sun, Douglas, Dessau, Hussain, and Shen}}]{Zhou:2005}
\bibinfo{author}{\bibfnamefont{X.}~\bibnamefont{Zhou}},
  \bibinfo{author}{\bibfnamefont{B.}~\bibnamefont{Wannberg}},
  \bibinfo{author}{\bibfnamefont{W.}~\bibnamefont{Yang}},
  \bibinfo{author}{\bibfnamefont{V.}~\bibnamefont{Brouet}},
  \bibinfo{author}{\bibfnamefont{Z.}~\bibnamefont{Sun}},
  \bibinfo{author}{\bibfnamefont{J.}~\bibnamefont{Douglas}},
  \bibinfo{author}{\bibfnamefont{D.}~\bibnamefont{Dessau}},
  \bibinfo{author}{\bibfnamefont{Z.}~\bibnamefont{Hussain}}, \bibnamefont{and}
  \bibinfo{author}{\bibfnamefont{Z.-X.} \bibnamefont{Shen}},
  \bibinfo{journal}{Journal of Electron Spectroscopy and Related Phenomena}
  \textbf{\bibinfo{volume}{142}}, \bibinfo{pages}{27 } (\bibinfo{year}{2005}),
  ISSN \bibinfo{issn}{0368-2048},
  \urlprefix\url{http://www.sciencedirect.com/science/article/pii/S0368204804003421}.

\bibitem[{\citenamefont{Hellmann et~al.}(2009)\citenamefont{Hellmann,
  Rossnagel, Marczynski-B\"uhlow, and Kipp}}]{Hell:2009}
\bibinfo{author}{\bibfnamefont{S.}~\bibnamefont{Hellmann}},
  \bibinfo{author}{\bibfnamefont{K.}~\bibnamefont{Rossnagel}},
  \bibinfo{author}{\bibfnamefont{M.}~\bibnamefont{Marczynski-B\"uhlow}},
  \bibnamefont{and} \bibinfo{author}{\bibfnamefont{L.}~\bibnamefont{Kipp}},
  \bibinfo{journal}{Phys. Rev. B} \textbf{\bibinfo{volume}{79}},
  \bibinfo{pages}{035402} (\bibinfo{year}{2009}),
  \urlprefix\url{http://link.aps.org/doi/10.1103/PhysRevB.79.035402}.

\bibitem[{\citenamefont{Passlack et~al.}(2006)\citenamefont{Passlack, Mathias,
  Andreyev, Mittnacht, Aeschlimann, and Bauer}}]{Passlack:2006}
\bibinfo{author}{\bibfnamefont{S.}~\bibnamefont{Passlack}},
  \bibinfo{author}{\bibfnamefont{S.}~\bibnamefont{Mathias}},
  \bibinfo{author}{\bibfnamefont{O.}~\bibnamefont{Andreyev}},
  \bibinfo{author}{\bibfnamefont{D.}~\bibnamefont{Mittnacht}},
  \bibinfo{author}{\bibfnamefont{M.}~\bibnamefont{Aeschlimann}},
  \bibnamefont{and} \bibinfo{author}{\bibfnamefont{M.}~\bibnamefont{Bauer}},
  \bibinfo{journal}{Journal of Applied Physics} \textbf{\bibinfo{volume}{100}},
  \bibinfo{eid}{024912} (\bibinfo{year}{2006}),
  \urlprefix\url{http://scitation.aip.org/content/aip/journal/jap/100/2/10.1063/1.2217985}.

\bibitem[{\citenamefont{Oloff et~al.}(2014)\citenamefont{Oloff, Oura,
  Rossnagel, Chainani, Matsunami, Eguchi, Kiss, Nakatani, Yamaguchi, Miyawaki
  et~al.}}]{Oloff:2014}
\bibinfo{author}{\bibfnamefont{L.-P.} \bibnamefont{Oloff}},
  \bibinfo{author}{\bibfnamefont{M.}~\bibnamefont{Oura}},
  \bibinfo{author}{\bibfnamefont{K.}~\bibnamefont{Rossnagel}},
  \bibinfo{author}{\bibfnamefont{A.}~\bibnamefont{Chainani}},
  \bibinfo{author}{\bibfnamefont{M.}~\bibnamefont{Matsunami}},
  \bibinfo{author}{\bibfnamefont{R.}~\bibnamefont{Eguchi}},
  \bibinfo{author}{\bibfnamefont{T.}~\bibnamefont{Kiss}},
  \bibinfo{author}{\bibfnamefont{Y.}~\bibnamefont{Nakatani}},
  \bibinfo{author}{\bibfnamefont{T.}~\bibnamefont{Yamaguchi}},
  \bibinfo{author}{\bibfnamefont{J.}~\bibnamefont{Miyawaki}},
  \bibnamefont{et~al.}, \bibinfo{journal}{New Journal of Physics}
  \textbf{\bibinfo{volume}{16}}, \bibinfo{pages}{123045}
  (\bibinfo{year}{2014}),
  \urlprefix\url{http://stacks.iop.org/1367-2630/16/i=12/a=123045}.

\bibitem[{\citenamefont{Damascelli et~al.}(1996)\citenamefont{Damascelli,
  Gabetta, Lumachi, Fini, and Parmigiani}}]{Damascelli:1996}
\bibinfo{author}{\bibfnamefont{A.}~\bibnamefont{Damascelli}},
  \bibinfo{author}{\bibfnamefont{G.}~\bibnamefont{Gabetta}},
  \bibinfo{author}{\bibfnamefont{A.}~\bibnamefont{Lumachi}},
  \bibinfo{author}{\bibfnamefont{L.}~\bibnamefont{Fini}}, \bibnamefont{and}
  \bibinfo{author}{\bibfnamefont{F.}~\bibnamefont{Parmigiani}},
  \bibinfo{journal}{Phys. Rev. B} \textbf{\bibinfo{volume}{54}},
  \bibinfo{pages}{6031} (\bibinfo{year}{1996}),
  \urlprefix\url{http://link.aps.org/doi/10.1103/PhysRevB.54.6031}.

\bibitem[{\citenamefont{i.~Tanaka}(2012)}]{Tanaka:2012a}
\bibinfo{author}{\bibfnamefont{S.}~\bibnamefont{i.~Tanaka}},
  \bibinfo{journal}{Journal of Electron Spectroscopy and Related Phenomena}
  \textbf{\bibinfo{volume}{185}}, \bibinfo{pages}{152 } (\bibinfo{year}{2012}),
  ISSN \bibinfo{issn}{0368-2048},
  \urlprefix\url{http://www.sciencedirect.com/science/article/pii/S0368204812000588}.

\bibitem[{\citenamefont{Yang et~al.}(2014)\citenamefont{Yang, Sobota,
  Kirchmann, and Shen}}]{Yang:2014aa}
\bibinfo{author}{\bibfnamefont{S.-L.} \bibnamefont{Yang}},
  \bibinfo{author}{\bibfnamefont{J.}~\bibnamefont{Sobota}},
  \bibinfo{author}{\bibfnamefont{P.}~\bibnamefont{Kirchmann}},
  \bibnamefont{and} \bibinfo{author}{\bibfnamefont{Z.-X.} \bibnamefont{Shen}},
  \bibinfo{journal}{Applied Physics A} \textbf{\bibinfo{volume}{116}},
  \bibinfo{pages}{85} (\bibinfo{year}{2014}), ISSN \bibinfo{issn}{0947-8396},
  \urlprefix\url{http://dx.doi.org/10.1007/s00339-013-8154-9}.

\bibitem[{\citenamefont{Riedl et~al.}(2009)\citenamefont{Riedl, Coletti,
  Iwasaki, Zakharov, and Starke}}]{Riedl:2009}
\bibinfo{author}{\bibfnamefont{C.}~\bibnamefont{Riedl}},
  \bibinfo{author}{\bibfnamefont{C.}~\bibnamefont{Coletti}},
  \bibinfo{author}{\bibfnamefont{T.}~\bibnamefont{Iwasaki}},
  \bibinfo{author}{\bibfnamefont{A.~A.} \bibnamefont{Zakharov}},
  \bibnamefont{and} \bibinfo{author}{\bibfnamefont{U.}~\bibnamefont{Starke}},
  \bibinfo{journal}{Physical Review Letters} \textbf{\bibinfo{volume}{103}},
  \bibinfo{eid}{246804} (pages~\bibinfo{numpages}{4}) (\bibinfo{year}{2009}),
  \urlprefix\url{http://link.aps.org/abstract/PRL/v103/e246804}.

\bibitem[{\citenamefont{Speck et~al.}(2011)\citenamefont{Speck, Jobst, Fromm,
  Ostler, Waldmann, Hundhausen, Weber, and Seyller}}]{Speck:2011}
\bibinfo{author}{\bibfnamefont{F.}~\bibnamefont{Speck}},
  \bibinfo{author}{\bibfnamefont{J.}~\bibnamefont{Jobst}},
  \bibinfo{author}{\bibfnamefont{F.}~\bibnamefont{Fromm}},
  \bibinfo{author}{\bibfnamefont{M.}~\bibnamefont{Ostler}},
  \bibinfo{author}{\bibfnamefont{D.}~\bibnamefont{Waldmann}},
  \bibinfo{author}{\bibfnamefont{M.}~\bibnamefont{Hundhausen}},
  \bibinfo{author}{\bibfnamefont{H.~B.} \bibnamefont{Weber}}, \bibnamefont{and}
  \bibinfo{author}{\bibfnamefont{T.}~\bibnamefont{Seyller}},
  \bibinfo{journal}{Applied Physics Letters} \textbf{\bibinfo{volume}{99}},
  \bibinfo{eid}{122106} (pages~\bibinfo{numpages}{3}) (\bibinfo{year}{2011}),
  \urlprefix\url{http://link.aip.org/link/?APL/99/122106/1}.

\bibitem[{\citenamefont{Barreto et~al.}(2013)\citenamefont{Barreto, Perkins,
  Johannsen, Ulstrup, Fromm, Raidel, Seyller, and Hofmann}}]{Barreto:2013}
\bibinfo{author}{\bibfnamefont{L.}~\bibnamefont{Barreto}},
  \bibinfo{author}{\bibfnamefont{E.}~\bibnamefont{Perkins}},
  \bibinfo{author}{\bibfnamefont{J.}~\bibnamefont{Johannsen}},
  \bibinfo{author}{\bibfnamefont{S.}~\bibnamefont{Ulstrup}},
  \bibinfo{author}{\bibfnamefont{F.}~\bibnamefont{Fromm}},
  \bibinfo{author}{\bibfnamefont{C.}~\bibnamefont{Raidel}},
  \bibinfo{author}{\bibfnamefont{T.}~\bibnamefont{Seyller}}, \bibnamefont{and}
  \bibinfo{author}{\bibfnamefont{P.}~\bibnamefont{Hofmann}},
  \bibinfo{journal}{Applied Physics Letters} \textbf{\bibinfo{volume}{102}},
  \bibinfo{eid}{033110} (pages~\bibinfo{numpages}{3}) (\bibinfo{year}{2013}),
  \urlprefix\url{http://link.aip.org/link/?APL/102/033110/1}.

\bibitem[{\citenamefont{Johannsen
  et~al.}(2013{\natexlab{b}})\citenamefont{Johannsen, Ulstrup, Bianchi, Hatch,
  Guan, Mazzola, Hornek{\ae}r, Fromm, Raidel, Seyller et~al.}}]{Johannsen:2013}
\bibinfo{author}{\bibfnamefont{J.~C.} \bibnamefont{Johannsen}},
  \bibinfo{author}{\bibfnamefont{S.}~\bibnamefont{Ulstrup}},
  \bibinfo{author}{\bibfnamefont{M.}~\bibnamefont{Bianchi}},
  \bibinfo{author}{\bibfnamefont{R.}~\bibnamefont{Hatch}},
  \bibinfo{author}{\bibfnamefont{D.}~\bibnamefont{Guan}},
  \bibinfo{author}{\bibfnamefont{F.}~\bibnamefont{Mazzola}},
  \bibinfo{author}{\bibfnamefont{L.}~\bibnamefont{Hornek{\ae}r}},
  \bibinfo{author}{\bibfnamefont{F.}~\bibnamefont{Fromm}},
  \bibinfo{author}{\bibfnamefont{C.}~\bibnamefont{Raidel}},
  \bibinfo{author}{\bibfnamefont{T.}~\bibnamefont{Seyller}},
  \bibnamefont{et~al.}, \bibinfo{journal}{Journal of Physics: Condensed Matter}
  \textbf{\bibinfo{volume}{25}}, \bibinfo{pages}{094001}
  (\bibinfo{year}{2013}{\natexlab{b}}),
  \urlprefix\url{http://stacks.iop.org/0953-8984/25/i=9/a=094001}.

\bibitem[{\citenamefont{Bostwick et~al.}(2010)\citenamefont{Bostwick, Speck,
  Seyller, Horn, Polini, Asgari, MacDonald, and Rotenberg}}]{Bostwick:2010}
\bibinfo{author}{\bibfnamefont{A.}~\bibnamefont{Bostwick}},
  \bibinfo{author}{\bibfnamefont{F.}~\bibnamefont{Speck}},
  \bibinfo{author}{\bibfnamefont{T.}~\bibnamefont{Seyller}},
  \bibinfo{author}{\bibfnamefont{K.}~\bibnamefont{Horn}},
  \bibinfo{author}{\bibfnamefont{M.}~\bibnamefont{Polini}},
  \bibinfo{author}{\bibfnamefont{R.}~\bibnamefont{Asgari}},
  \bibinfo{author}{\bibfnamefont{A.~H.} \bibnamefont{MacDonald}},
  \bibnamefont{and}
  \bibinfo{author}{\bibfnamefont{E.}~\bibnamefont{Rotenberg}},
  \bibinfo{journal}{Science} \textbf{\bibinfo{volume}{328}},
  \bibinfo{pages}{999} (\bibinfo{year}{2010}),
  \urlprefix\url{http://www.sciencemag.org/cgi/content/abstract/328/5981/999}.

\bibitem[{\citenamefont{Ristein et~al.}(2012)\citenamefont{Ristein, Mammadov,
  and Seyller}}]{Ristein:2012}
\bibinfo{author}{\bibfnamefont{J.}~\bibnamefont{Ristein}},
  \bibinfo{author}{\bibfnamefont{S.}~\bibnamefont{Mammadov}}, \bibnamefont{and}
  \bibinfo{author}{\bibfnamefont{T.}~\bibnamefont{Seyller}},
  \bibinfo{journal}{Phys. Rev. Lett.} \textbf{\bibinfo{volume}{108}},
  \bibinfo{pages}{246104} (\bibinfo{year}{2012}),
  \urlprefix\url{http://link.aps.org/doi/10.1103/PhysRevLett.108.246104}.

\bibitem[{\citenamefont{Cacho et~al.}(2014)\citenamefont{Cacho, Petersen,
  Gierz, Liu, Kaiser, Chapman, Turcu, Cavalleri, and Springate}}]{Cacho:2014}
\bibinfo{author}{\bibfnamefont{C.}~\bibnamefont{Cacho}},
  \bibinfo{author}{\bibfnamefont{J.}~\bibnamefont{Petersen}},
  \bibinfo{author}{\bibfnamefont{I.}~\bibnamefont{Gierz}},
  \bibinfo{author}{\bibfnamefont{H.}~\bibnamefont{Liu}},
  \bibinfo{author}{\bibfnamefont{S.}~\bibnamefont{Kaiser}},
  \bibinfo{author}{\bibfnamefont{R.}~\bibnamefont{Chapman}},
  \bibinfo{author}{\bibfnamefont{E.}~\bibnamefont{Turcu}},
  \bibinfo{author}{\bibfnamefont{A.}~\bibnamefont{Cavalleri}},
  \bibnamefont{and}
  \bibinfo{author}{\bibfnamefont{E.}~\bibnamefont{Springate}}, in
  \emph{\bibinfo{booktitle}{19th International Conference on Ultrafast
  Phenomena}} (\bibinfo{publisher}{Optical Society of America},
  \bibinfo{year}{2014}),
  \urlprefix\url{http://www.opticsinfobase.org/abstract.cfm?URI=UP-2014-07.Mon.P1.35}.

\bibitem[{\citenamefont{Frassetto et~al.}(2011)\citenamefont{Frassetto, Cacho,
  Froud, Turcu, Villoresi, Bryan, Springate, and Poletto}}]{Frassetto:2011}
\bibinfo{author}{\bibfnamefont{F.}~\bibnamefont{Frassetto}},
  \bibinfo{author}{\bibfnamefont{C.}~\bibnamefont{Cacho}},
  \bibinfo{author}{\bibfnamefont{C.~A.} \bibnamefont{Froud}},
  \bibinfo{author}{\bibfnamefont{I.~C.~E.} \bibnamefont{Turcu}},
  \bibinfo{author}{\bibfnamefont{P.}~\bibnamefont{Villoresi}},
  \bibinfo{author}{\bibfnamefont{W.~A.} \bibnamefont{Bryan}},
  \bibinfo{author}{\bibfnamefont{E.}~\bibnamefont{Springate}},
  \bibnamefont{and} \bibinfo{author}{\bibfnamefont{L.}~\bibnamefont{Poletto}},
  \bibinfo{journal}{Opt. Express} \textbf{\bibinfo{volume}{19}},
  \bibinfo{pages}{19169} (\bibinfo{year}{2011}),
  \urlprefix\url{http://www.opticsexpress.org/abstract.cfm?URI=oe-19-20-19169}.

\bibitem[{\citenamefont{Ulstrup
  et~al.}(2014{\natexlab{b}})\citenamefont{Ulstrup, Johannsen, Grioni, and
  Hofmann}}]{ulstrupc:2014}
\bibinfo{author}{\bibfnamefont{S.}~\bibnamefont{Ulstrup}},
  \bibinfo{author}{\bibfnamefont{J.~C.} \bibnamefont{Johannsen}},
  \bibinfo{author}{\bibfnamefont{M.}~\bibnamefont{Grioni}}, \bibnamefont{and}
  \bibinfo{author}{\bibfnamefont{P.}~\bibnamefont{Hofmann}},
  \bibinfo{journal}{Review of Scientific Instruments}
  \textbf{\bibinfo{volume}{85}}, \bibinfo{eid}{013907}
  (\bibinfo{year}{2014}{\natexlab{b}}),
  \urlprefix\url{http://scitation.aip.org/content/aip/journal/rsi/85/1/10.1063/1.4863322}.

\end{thebibliography}

\end{document}